\documentclass[aps,prl,twocolumn,superscriptaddress,showpacs]{revtex4}  % for review and submission
\usepackage{graphicx}  % needed for figures
\usepackage{dcolumn}   % needed for some tables
\usepackage{bm}        % for math
\usepackage{amssymb}   % for math
\usepackage[utf8x]{inputenc}

\begin{document}

\title{Quantum levitation of nanoparticles seen with ultracold neutrons}

%\date{\today}

\author{V.V. Nesvizhevsky}
\affiliation{Institut Laue-Langevin, 6, rue Jules Horowitz, F-38042 Grenoble, France}
\author{A.Yu. Voronin}
\affiliation{Lebedev Institute, 53 Leninsky pr., RU-119991 Moscow,
Russia}
\author{A. Lambrecht}
\affiliation{Laboratoire Kastler-Brossel, CNRS, ENS, UPMC, Case 74,
F-75252 Paris, France}
\author{ S. Reynaud}
\affiliation{Laboratoire Kastler-Brossel, CNRS, ENS, UPMC, Case 74,
F-75252 Paris, France}
\author{E.V. Lychagin}
\affiliation{Joint Institute for Nuclear Research, 6 Joliot Curie, RU-141980 Dubna, Russia}
\author{A.Yu. Muzychka}
\affiliation{Joint Institute for Nuclear Research, 6 Joliot Curie, RU-141980 Dubna, Russia}
\author{A.V. Strelkov}
\affiliation{Joint Institute for Nuclear Research,  6 Joliot Curie, RU-141980 Dubna, Russia}
\pacs{ }

\begin{abstract}
Analyzing new experiments with ultracold neutrons (UCNs) we show
that physical adsorption of nanoparticles/nano-droplets, levitating
in high-excited states in a deep and broad potential well formed by
van der Waals/Casimir-Polder (vdW/CP) forces results in new effects
on a cross-road of fundamental interactions, neutron, surface and
nanoparticle physics. Accounting for the interaction of UCNs with
nanoparticles explains a recently discovered intriguing small
heating of UCNs in traps. It might be relevant to the striking
conflict of the neutron lifetime experiments with smallest reported
uncertainties by adding false effects there.
\end{abstract}

\maketitle

Surface diffusion of atoms, molecules and clusters, physically
adsorbed in the potential well associated with van der Waals and
Casimir-Polder interactions (vdW/CP) plays a key role in various
phenomena in physics, chemistry, biology, and in applications
\cite{1,2,3,4}. The interaction is affected at small distance by
"close-to-contact" effects depending on roughness and surface state
\cite{5,6}. Because of experimental limitations the precision
studies are usually restricted to probes having either molecular or
macroscopic sizes. Here we address the less extensively explored
field of physical adsorption of nanoparticles and discover
qualitatively new universal phenomena related to their sizes and
masses. Rigorous theoretical formalism describing states of
physically adsorbed nanoparticles and their interaction with
ultracold neutrons (UCN) \cite{7,8,9} is given in a longer article
\cite{10}; however general features can be understood from arguments
stated below. The depth of the potential well affecting the motion
of a sufficiently small physisorbed particle near a surface is proportional to
the number of atoms in it, while the thermal energy attributed to
the particle is always $3/2 k_{\mathrm{B}}T$, where $k_{\mathrm{B}}$
is the Boltzmann constant and $T$ the temperature. Thus a
nanoparticle is found near the surface in a deep and broad potential
well in normal direction; nanoparticles in states with low quantum number $n$ are
strongly bound to the surface, while those in high-$n$ states move
along it, with the effects of roughness and inhomogeneities mixing
the two velocity components. Nanoparticles in high-$n$ states might
form a two-dimensional "cloud". The number of states is so large
that a nanoparticle exhibits typically a quasi-classical motion
fitting the conditions searched for in \cite{11} for solving puzzles
in UCN physics.

Here, we report on the theoretical justification for existence of
levitating nanoparticles, new treatment of the relevant existing UCN
experimental data, and the results of new dedicated experiments,
which provide evidence for levitating nanoparticles over solid and
liquid surfaces. We show that our theoretical model describes all
relevant existing and new UCN data; to our knowledge there are no
alternative models, which could describe them. After a few reminders
on the vdW/CP potential, we describe scattering of UCN on
nanoparticles, discuss experimental results and consider
consequences of our findings.

The shape of the potential is calculated using a general expression
involving only the scattering properties of the two objects
\cite{12}. For example, the case of diamond nanospheres above a
copper plane has been studied in great details in \cite{13}.
Electromagnetic waves scattering on a plate is characterized by
Fresnel reflection amplitudes, the form of which is fixed by the
dielectric response function $\varepsilon$ of copper. For
frequencies $\omega$ lower than the plasma frequency
$\omega_{\mathrm{P}}$, $\varepsilon$ is large and copper behaves as
a good reflector. For larger frequencies the reflection properties
are poorer. When considering a nanoparticle of vanishing radius, the
electric dipole approximation is sufficient. Calculations based on a
simple model for the dielectric function of diamond containing only
one resonance frequency $\omega_1$ reproduce the well-known vdW/CP
energy. For distances small compared to the plasma wavelength (136nm
for copper), or the resonance wavelength (106nm for diamond), the
interaction reduces to the commonly used vdW formula with a power
law $R^3/L^3$ in the vicinity of the surface, where $R$ is the
nanoparticle radius, and $L$ the distance of closest approach to the
plane.

For nanoparticles of arbitrary size the quantum dispersion energy
must be calculated by integrating the phaseshifts corresponding to
all modes of the electromagnetic vacuum. In particular, many Mie
scattering amplitudes (corresponding to many multipole components
beyond the electric dipole), contribute significantly to the effect
when the particle is close to the surface \cite{12}. The exact
solution for the interaction energy predicts a smoother power law
$R/L$  close to contact with the surface, thus leading to a regular
solution for the Schr\"odinger equation, in contrast to the commonly
used vdW formula.

We have analyzed the Schr\"odinger equation for the quantum
dispersion potential and will sketch our findings here; details of
the calculations are given in \cite{10}. We found a large number of
bound states with the eigenenergies $E_n$ deeper than $3/2
k_\mathrm{B}T$, localized at distances of up to a few nanometers to
the surface, provided $R\gtrsim 1$nm. The nanoparticle thus might
fly over the surface while levitating in high-$n$ quantum states.
More precise statements depend on the population and thermalization
dynamics and are sensitive to the low-$n$ quantum states spectrum.
In contrast to high-$n$ states, where the nanoparticles stay
essentially far from the surface, the precise properties of low-$n$
states depend on close-to-contact details and in particular on the
effect of roughness. Consequences of this effect can be
characterized by using the methods developed for studying the
Casimir force between two metallic plates. The probability of
approach to close distances is reduced by contact repulsion from the
highest peaks of the roughness profile \cite{5}, while the
dispersion force is affected by the distribution of approach
distances due to roughness \cite{6}. Here we chose a simple
parametric approach: on top of the attraction calculated in
\cite{6}, we introduce contact repulsion so that the effective
potential has a minimum at a distance equal to 1-3nm for the solid
samples used. Because of the large number of high-$n$ quantum states
for the bound nanoparticles, a quasi-classical description of their
motion in the vdW/CP potential is sufficient. Thus, when UCN bounce
on the surface covered by levitating nanoparticles, we essentially
deal with events of Doppler shift in UCN energy in the laboratory
frame due to their elastic (in the centre of mass reference)
collisions with nanoparticles in high-$n$ quantum states. Due to
extremely low temperature of UCN ($<1$mK), small energy ($<10^{-7}$eV),
and low velocity (a few m/s) UCNs possess the unique property of
total elastic reflection from motionless surfaces at any incidence
angle. The probability of UCN loss per one bounce might be quite low:
theoretically predicted probabilities of losses can reach $\sim 10^{-9}$;
the best experimentally achieved values are $\sim 10^{-6}$. 
The dominant loss mechanisms are nuclear absorbtion and
up-scattering on phonons in the surface material. Even at the
ambient temperature the probability of up-scattering is much 
smaller than the probability of coherent elastic reflection. 
 
That is why UCNs can be stored in closed traps for extended
periods thus providing an extremely sensitive probe for rare
processes or weak interactions. When studying UCNs storage in traps,
unusual inelastic scattering of UCNs on trap surfaces was discovered
\cite{14,15,16,17,18}. One calls this phenomenon "small heating of
UCNs" and such up-scattered neutrons Vaporizing UCNs (VUCN) - in
analogy to vaporization of molecules. First measurements were
followed by studies of several research groups, which essentially
confirmed the initial observation, but suggested controversial
estimations of the VUCNs production rates.

Let us first give strong experimental motivation for the existence
of levitating nanoparticles above surfaces. In the following we will
show VUCN spectra that we calculated with a method sketched below
using the initial UCN spectrum and the spectrometer spectral
efficiency that we had measured experimentally. It has been found in
\cite{18} that the count rate of VUCN produced on stainless steel
surfaces as a function of the sample pre-heating temperature shows a
sharp increase, the so-called "temperature resonance" shown in the
left part of Fig.1.
\begin{figure}[h]
\centering
\includegraphics[width=9cm]{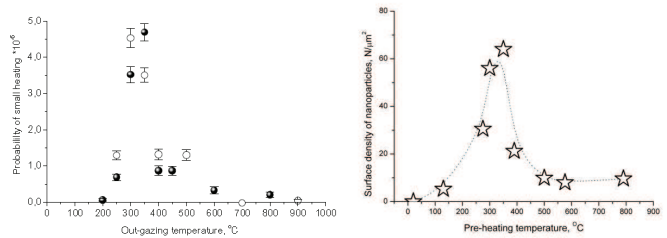}
\caption{(left) The probability of small heating of UCNs on surface
of stainless steel samples is shown as a function of the temperature
of sample outgasing (pre-heating); the measurement is performed at
300K. Black and white points show results of two independent
experiments using analogous samples. (right) The surface density of
nanostructures with a radius of 6-7nm (stars) corresponding to the
resonance enhancement in the efficiency of VUCN detection, observed
in AFM. More details in the body text.} \label{Fig1}
\end{figure}
On the right part of Fig.1 we show the surface density of
nanostructures as a function of temperature measured with an Atomic
Force Microscope (AFM). These measurements indicate intense
formation of nanostructures on the surface which takes place
precisely at the temperature of sharp increase in the VUCN count
rate. Nanostructures' size increases monotonously as a function of
the heating temperature: a few particles at the ambient temperature,
intense growth of the number and size of nanostructures below ~350C;
further increase in size but decrease in number of nanoparticles
above ~350C (some nanoparticles coagulate). While no alternative
explanation has yet been proposed, we regard this as indication that it is
due to levitating nanoparticles.

Let us now briefly describe the basic equations of our model. The
average differential cross-section of the interaction of an UCN with
a nanoparticle is \cite{10}:
\begin{eqnarray}
&&\frac{\mathrm{d}\overline{\sigma}}{\mathrm{d}E}=\frac{1}{2mk_n}\left(\frac{M}{2\pi
k_{\mathrm{B}} T}\right)^{3/2} \times \\
&&\int_0^{\infty} \mathrm{d}V \exp\left(-\frac{MV^2}{2k_{\mathrm{B}}
T}\right)
\int_{k_{\mathrm{min}}}^{k_{\mathrm{max}}}\mathrm{d}k_0\int_0^{2\pi}
\mathrm{d}\varphi |f_{\mathrm{B}}(k_0,V,E,\varphi)|^2 \nonumber
\label{crosssection}
\end{eqnarray}
with $m$, $v_n$, $k_n=mv_n$ the neutron mass, initial velocity and
momentum and $M$, $V$ the nanoparticle mass and velocity. The
kinematically allowed area of integration
$[k_{\mathrm{min}},k_{\mathrm{max}}]$ is determined by the
conditions
\begin{eqnarray}
|m V-k_n| \le k_0 \le m V +k_n \nonumber \\
|\sqrt{k_n^2+2mE}-m V| \le k_0 \le \sqrt{k_n^2+2mE}+m V
\label{condition}
\end{eqnarray}
$E$ is the energy transfer and
$k_0=m|\overrightarrow{v}_n-\overrightarrow{V}| $ the incident
momentum modulus in the center-of-mass reference system. 
The Born amplitude $f_{\mathrm{B}}$ of
the neutron nanoparticle scattering is
$f_{\mathrm{B}}(q)=-2mRU_0/q^2(\hbar
\sin(qR/\hbar)/(qR)-\cos(qR/\hbar))$, where $q$ is the transferred
momentum, and $U_0$ is the nanoparticle-neutron optical potential.
To estimate relevant nanoparticle radii, we use the above formalism
and follow the general approach from \cite{11}: 1) $R$ should not be
too small, otherwise the coherent interaction cross-section is too
low (as $\mathrm{d}\overline{\sigma}/\mathrm{d}E \sim R^6$), and 2)
$R$ should not be too large, otherwise nanoparticles are too heavy;
thus their thermal velocity is too small, $q$ is too small, and
VUCNs could not be found in the window of efficient VUCN detection.

We first apply our method to the small heating of UCNs on solid
surfaces, where we have estimated in several manners the effective
radius $\overline{R}$ of nanoparticles contributing to the observed
inelastic process. The first sample \cite{16,17} we consider is
powder of diamond nanoparticles with radii of 1-10 nm, on a copper
surface, or above "sand" of diamond nanoparticles. We estimate the
effective radius by the following three methods. 1) Using measured
shapes of the differential VUCN spectra leading to
$\overline{R}^{\mathrm{spectrum}}_{\mathrm{diamond}}=9.4 \pm 0.4$nm.
2) Using measured temperature dependencies of the count rates of
detected VUCNs which gives
$\overline{R}^{\mathrm{temp}}_{\mathrm{diamond}}=9.5 \pm 0.6$nm; 3)
Using the measured size distribution of nanoparticles:
$\overline{R}^{\mathrm{aver. size}}_{\mathrm{diamond}}=8.5 \pm
0.5$nm. Agreement between these independent estimations is
convincing keeping in mind that they are not very sensitive to such
parameters as the nanoparticle size and shape distribution,
clustering, surface roughness, impurities etc. In contrast, if one
abandons the "free levitation" model and "tunes" one critical
parameter alone, say $\overline{R}$ by 10-20\%, theoretical
predictions largely contradict the data. Strictly speaking, there is
no reason then, why all these estimations of $\overline{R}$ are
equal, or even comparable. Another measured sample is nanoparticles
appearing on stainless steel surface due its thermal treatment
\cite{17,18}. Again, the two independent estimations of the
effective radii
$\overline{R}^{\mathrm{spectrum}}_{\mathrm{steel}}=6.3 \pm 0.3$nm
and $\overline{R}^{\mathrm{aver. size}}_{\mathrm{steel}}=6.6 \pm
0.3$nm agree. The effective masses of diamond and stainless steel
nanoparticles estimated above are equal, thus providing an
additional test of validity of our model.

Small heating of UCN on liquid Fomblin oils has been observed as
well \cite{14,15,16} but no detailed spectral measurements have been
performed. In view of the above evidence we assume that this
phenomenon might be due to collisions of UCNs with levitating
Fomblin oil nano-droplets. This mechanism would complement the known
effects due to surface capillary waves \cite{19} and surface thermal
fluctuations \cite{20}. We mention but not analyze here other
hypotheses on origin of VUCN \cite{21,22,23}; as they do not provide
quantitative predictions and/or do conflict with experiments. While
solid nanoparticles are characterized by long formation and
evolution times, the number and size of levitating liquid
nano-droplets \cite{31,32}    
could rapidly change with temperature, or other
conditions like pressure, as they are governed by equilibrium of
permanent formation of nano-droplets from vapors, also from
explosion of bubbles on surface, and of their evaporation.

To compare these models, we measured VUCN count rates as a function
of temperature, shown in Fig.2. Although our model does not describe
explicitly the number of nano-droplets as a function of temperature,
it is natural to assume that this number decreases with falling temperature
as the Fomblin oil vapor pressure above surface, in analogy to the
present data. Alternative interpretations of neutron small-heating,
such as the hypothesis of thermal fluctuations of the surface
\cite{20}, and the hypothesis of capillary surface waves \cite{10},
seem to contradict these experimental results, as they propose
either too low absolute probabilities \cite{20}, or/and inverse
temperature dependence of the probability \cite{19}. In contrast our
model gives the correct qualitative behavior.
\begin{figure}[h]
\centering
\includegraphics[width=9cm]{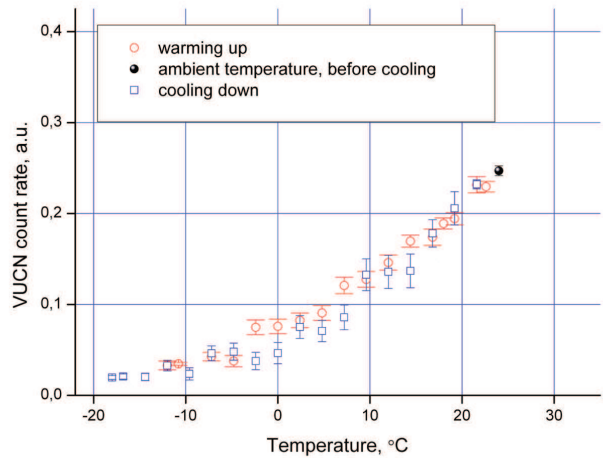}
\caption{VUCN count rate in a.u. on a Fomblin surface as a function
of temperature. The circle indicates the data at ambient temperature
(the total probability of small heating is $10^{-5}$ per wall
collision); open squares show data measured when cooling the sample
down, and open circles correspond to data measured when warming the
sample up. } \label{Fig2}
\end{figure}

In order to assess this question more precisely we have calculated
the VUCN spectrum within our model and compared it to results of a
new dedicated precision study of small heating of UCNs on solid and
liquid surfaces in our precisely calibrated spectrometer BGS shown
in Fig.3.
\begin{figure}[h]
\centering
\includegraphics[width=7cm]{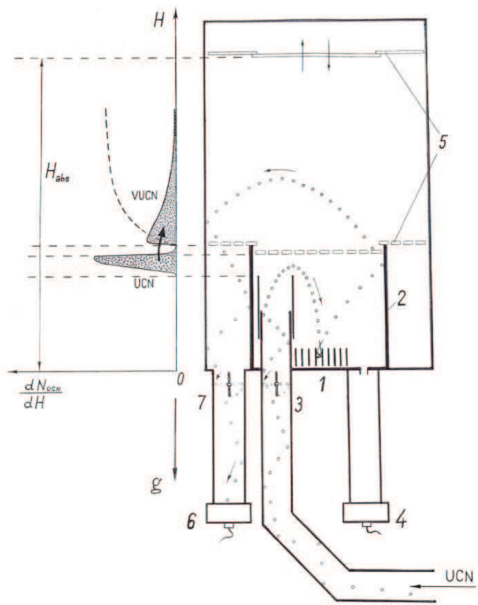}
\caption{Layout of the Big Gravitational Spectrometer (BGS): (1)
sample, (2) gravitational barrier, (3) entrance valve, (4) UCN
monitor detector, (5) UCN absorber, (6) VUCN detector, and (7) exit
valve. The principle of the procedure to measure small energy
transfers is sketched in the insert on the left side. From bottom to
top (in scale): initial differential UCN spectrum (the mean energy
is 31neV , the half-width of the spectral mono-line is 3neV), a
dead-zone of 3neV insensitive to VUCNs, the differential and
integral (dashed line) spectra of VUCN. The differential efficiency
of VUCN detection is calculated using precisely measured values of
UCN and VUCN storage times as a function of their energy; decrease
in the detection efficiency, caused by partial losses of neutrons in
samples, is taken into account.} \label{Fig3}
\end{figure}
The comparison is shown in Fig.4 for several different physical
systems, i.e. ultra-diamond ($\overline{R}\sim $2.5nm) and sapphire
($\overline{R}\sim $10nm) nanoparticles with broad size
distribution, thin (1 $\mu$m) and thick (a few mm) layers of Fomblin
oil, and naturally growing nanoparticles on a copper surface.
\begin{figure}[h]
\centering
\includegraphics[width=9cm]{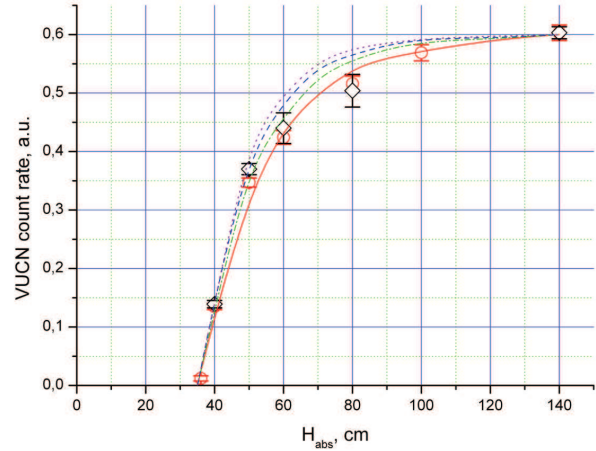}
\caption{VUCN count rate as a function of neutron energy, expressed
in UCN raising height in the Earth's gravitational field, in cm.
Circles show results measured with various solid nanoparticles:
diamond, sapphire, copper; all these results agree within
statistical accuracy. Rhombi indicate data measured with a Fomblin
oil sample; they agree with the data on solid nanoparticles. Four
lines correspond to our model calculations of VUCN spectra on the
Fomblin surface, for four hypotheses on the size distribution of
nano-droplets:$(R/R_0)^{-l}$ where $l=1,2,3,4$ from top to bottom.
The lines and the data are normalized to equal count rate at
infinite height.} \label{Fig4}
\end{figure}
Clearly all integral spectra are equivalent within statistical
accuracy showing that there is a universal behavior for these very
different physical systems. This is a natural consequence of our
model. In fact, the sensitivity of the spectrometer is sharply
shaped to some nanoparticle mass, which, in turn, depends on the
initial UCN energy range and the window of the spectrometer
sensitivity. However, within the peak resonance sensitivity, there
is a slight dependence on the mass distribution, which we show in
Fig.4. If the size distribution is known, the corresponding
uncertainties are small. Rotations, shape and non-uniformity of
nanoparticles, interference of scattering on neighbor nanoparticles
and consequent scattering on one nanoparticle have been disregarded
here. Concerning the results for Fomblin oil, they give 
evidence that the dominant mechanism of small heating of UCN on
Fomblin oil surface in our experiment is UCN scattering on
levitating nano-droplets.

The present study was motivated in part by the unsatisfactory status
of the neutron lifetime $\tau_n$ experiments which show large and so
far unexplained discrepancy between results with smallest reported
uncertainties \cite{24,25,26}. This contradiction has been discussed
with far-reaching consequences of an eventual shift of the mean
world value for fundamental particle physics and cosmology
\cite{27,28,29}, but only a single study has tried to verify
independently the validity of the measured results \cite{30}. All
these experiments use UCN traps with Fomblin-oil walls, and assume
conservation of UCN energy. However using our data and model, it is
easy to show that a major fraction of UCNs do change their energy,
so that their loss rate is different from the assumed values and
major false effects might arise. In particular, as the rate of
forming Fomblin-oil nano-droplets depends on experimental
conditions, the basic idea of these experiments is compromised: the
geometrical and energy methods of UCN loss extrapolation cannot be
applied without reservations. Nano-droplet production rates depend
on parameters never properly controlled like pre-history of the
Fomblin oil treatment or vacuum. In particular, the UCN loss
probability would be different in large and small traps; it would
also evolve in time. The task of estimating reliable corrections to
$\tau_n$  values goes beyond the scope of this article. An
attractive application might consist in studying vdW/CP interaction
between levitating nanoparticles and surfaces. Corresponding
corrections to the integral VUCN spectra account for $\sim10\%$ even
in the present study, not optimized for these purposes. Surface
potentials define the distribution of distances of levitating
nanoparticles to surface. As UCN could be up-scattered twice on one
nanoparticle, before reflection from the surface and afterwards, the
spectra would be modified depending on the distance to surface.
Besides, complementary to using trapped atomic clusters for
decorating surface defects, boundaries, step edges, grain
boundaries, elastic strain fields, UCNs are sensitive to
nanoparticles in motion above defect-free zones, thus giving us
access to nanoparticle mobility. In levitation, nanoparticle
mobility is very high; on the other hand chemical interactions
largely reduce it thus giving us access to chemical properties of
nanoparticles and surfaces. In an analogous way, one could also
measure electrostatic effects.

We thank our colleagues for useful discussions, in particular A.L.
Barabanov, S.T. Belyaev, H.G. Boerner, L.N. Bondarenko, D. Delande,
D. Dubbers, R. Golub, V.K. Ignatovich, A. Leadbetter, S.K.
Lamoreaux, V.I. Morozov, G.V. Nekhaev, S. Paul, J.M. Pendlebury, G.
Pignol, K.V. Protasov. The authors benefited from exchange of ideas
within the ESF Research Network CASIMIR, and the GRANIT
collaboration.

\end{document}